\documentclass{revtex4}

\usepackage{graphicx}

\input epsf

\begin{document}
\bibliographystyle{revtex}

%===================================================================================
\rightline{\tt hep-th/0507201}
\bigskip

\title{How to Extrapolate A String Model to Finite Temperature:\\
Interpolations and Implications for the Hagedorn Transition\\}

\author{Keith R. Dienes\footnote{E-mail address:  {\tt dienes@physics.arizona.edu}},
        Michael Lennek\footnote{E-mail address:  {\tt mlennek@physics.arizona.edu}}}
\affiliation{Department of Physics, University of Arizona, Tucson, AZ  85721  USA}

\date{July 20, 2005}

\begin{abstract}
     In this paper, we discuss the important question of how to extrapolate a
     given zero-temperature string model to finite temperature.  It turns out that
     this issue is surprisingly subtle, and we show that many of the standard
     results require modification.
     For concreteness, we focus on the case of the ten-dimensional $SO(32)$ heterotic string,
     and show that the usual finite-temperature extrapolation for this string
     is inconsistent at the level of a proper worldsheet theory.
     We then derive the proper extrapolation,
     and in the process uncover a universal Hagedorn temperature for 
     all tachyon-free closed string theories in ten dimensions --- 
     both Type~II and heterotic.
     As we discuss, these results are not in conflict with the 
     well-known exponential growth in the degeneracies of string states in such models.  
     This writeup is a concise summary of our recent 
     paper {\tt hep-th/0505233},
     here presented using a ``bottom-up'' approach 
     based on determining self-consistent finite-temperature extrapolations
     of zero-temperature string models.
     Some new results and observations are also added.
\end{abstract}

\maketitle

\def\beq{\begin{equation}}
\def\eeq{\end{equation}}
\def\beqn{\begin{eqnarray}}
\def\eeqn{\end{eqnarray}}
\def\half{{\textstyle{1\over 2}}}
\def\quarter{{\textstyle{1\over 4}}}

\def\calO{{\cal O}}
\def\calE{{\cal E}}
\def\calP{{\cal P}}
\def\calT{{\cal T}}
\def\calM{{\cal M}}
\def\calF{{\cal F}}
\def\calS{{\cal S}}
\def\calY{{\cal Y}}
\def\calV{{\cal V}}
\def\calZ{{\cal Z}}
\def\calN{{\cal N}}
\def\ibar{{\overline{\imath}}}
\def\chibar{{\overline{\chi}}}
\def\ttwo{{\vartheta_2}}
\def\tthree{{\vartheta_3}}
\def\tfour{{\vartheta_4}}
\def\ttwob{{\overline{\vartheta}_2}}
\def\tthreeb{{\overline{\vartheta}_3}}
\def\tfourb{{\overline{\vartheta}_4}}

\def\qbar{{\overline{q}}}
\def\mm{{\tilde m}}
\def\nn{{\tilde n}}
\def\rep#1{{\bf {#1}}}
\def\ie{{\it i.e.}\/}
\def\eg{{\it e.g.}\/}

\newcommand{\newc}{\newcommand}
\newc{\gsim}{\lower.7ex\hbox{$\;\stackrel{\textstyle>}{\sim}\;$}}
\newc{\lsim}{\lower.7ex\hbox{$\;\stackrel{\textstyle<}{\sim}\;$}}

%==============================================================================
\hyphenation{su-per-sym-met-ric non-su-per-sym-met-ric}
\hyphenation{space-time-super-sym-met-ric}
\hyphenation{mod-u-lar mod-u-lar--in-var-i-ant}
%==============================================================================

%================== BLACKBOARD BOLD CHARACTERS ==============================

\def\inbar{\,\vrule height1.5ex width.4pt depth0pt}

\def\IC{\relax\hbox{$\inbar\kern-.3em{\rm C}$}}
\def\IQ{\relax\hbox{$\inbar\kern-.3em{\rm Q}$}}
\def\IR{\relax{\rm I\kern-.18em R}}
 \font\cmss=cmss10 \font\cmsss=cmss10 at 7pt
\def\IZ{\relax\ifmmode\mathchoice
 {\hbox{\cmss Z\kern-.4em Z}}{\hbox{\cmss Z\kern-.4em Z}}
 {\lower.9pt\hbox{\cmsss Z\kern-.4em Z}}
 {\lower1.2pt\hbox{\cmsss Z\kern-.4em Z}}\else{\cmss Z\kern-.4em Z}\fi}

% Redefine caption to put text and formulas in smaller font
\long\def\@caption#1[#2]#3{\par\addcontentsline{\csname
  ext@#1\endcsname}{#1}{\protect\numberline{\csname
  the#1\endcsname}{\ignorespaces #2}}\begingroup
    \small
    \@parboxrestore
    \@makecaption{\csname fnum@#1\endcsname}{\ignorespaces #3}\par
  \endgroup}
\catcode`@=12

\input epsf
%============================== TEXT BEGINS HERE ============================

%=============================================================================
\section{Preliminaries:  ~~Thermal orbifolds and partition functions}
\setcounter{footnote}{0}

Our first task is to understand how to extrapolate a 
given zero-temperature theory to finite temperature.
In this section, we shall quickly review the standard procedures for doing this.

Within the language of quantum field theory, there is a well-known procedure 
for extrapolating
a given zero-temperature model to finite temperature $T$:  
we compactify a Euclidean time-like dimension
on a circle of radius $R_T = (2\pi T)^{-1}$, and require
that bosonic (fermionic) fields be periodic (anti-periodic) around this circle.
The thermal Matsubara modes are then
nothing but the Kaluza-Klein states corresponding to this compactification,
and the periodic (anti-periodic) boundary conditions for bosons (fermions)
are imposed in order to ensure proper thermal spin-statistics relations.

This basic picture also holds in string theory~\cite{Polbook,Pol86}:
again we compactify a Euclidean time-like dimension on a circle of radius $R_T$,
obtaining a theory with one fewer spacetime dimension,
and again we demand appropriate boundary conditions for spacetime bosons/fermions
around the thermal circle.
However, for closed strings, there is a new feature:  in addition to Kaluza-Klein
(Matsubara) momentum states, there are also Matsubara winding states which
arise from closed strings wrapping around this thermal circle.
Such states are needed for modular invariance, and encourage us to think
of this thermal circle quite literally as a compactified spacetime dimension.

It will be important for us to see how this works at the level 
of explicit string model-building and the associated thermal partition functions.
The following discussion follows the mathematical treatment in Ref.~\cite{Rohm}, 
suitably T-dualized in order to apply to temperature rather 
than geometric radius~\cite{AtickWitten}.  
Let us suppose that we begin with a $D$-dimensional zero-temperature closed string model 
whose one-loop partition function is given by $Z(\tau)$,
where $\tau$ is the complex toroidal modular parameter. 
The first step in the thermal construction 
is to compactify this theory on circle of radius $R_T$.
At this stage, we then have a thermal 
string partition function $Z_{\rm therm}(\tau,T)$ 
of the form
\beq
        Z_{\rm therm}(\tau,T) ~\equiv~ Z(\tau) \, Z_{\rm circ}(\tau,T)~
\label{parfunct}
\eeq
where the extra factor $Z_{\rm circ}$
represents a double summation over integer Matsubara momentum and winding modes:
\beq
     Z_{\rm circ}(\tau,T)~=~
     \sqrt{ \tau_2}\,
    \sum_{m,n\in\IZ} \,
      \overline{q}^{(ma-n/a)^2/4}  \,q^{(ma+n/a)^2/4}
\label{Zcircdef}
\eeq
with $a\equiv 2\pi T/M_{\rm string}$ and $\tau_2\equiv {\rm Im}\,\tau$.
Here $(m,n)$ represent the thermal momentum and winding numbers, respectively.
However, at this stage in the construction, we see that each of the states within
$Z(\tau)$ is multiplied by the same thermal spectrum of 
integer momentum and winding modes within $Z_{\rm circ}$.  The next step, therefore,
is to break this degeneracy, ensuring that while bosonic states within 
$Z(\tau)$ continue to have integer Matsubara modes, fermionic states
should have {\it half-integer}\/ modings (so that they are anti-periodic
around the thermal circle).  

In string theory, the only way to accomplish this in a self-consistent manner
is by twisting or orbifolding the compactified theory 
in Eq.~(\ref{parfunct}).  What orbifold do we choose?
Clearly, we need a $\IZ_2$ operator that distinguishes between 
spacetime bosonic and fermionic states, such as $(-1)^F$ (where $F$ represents
spacetime fermion number).  We shall generally let $Q$ denote such an operator, since
we shall eventually argue that $Q$ must contain more than merely $(-1)^F$.  
However, we will also need to couple $Q$ with an operator that can distinguish
between between integer and half-integer thermal momenta.
As we shall see,
such an operator is given by $\calT:  y\to y+\pi R_T$, where $y$
is the (T-dual) coordinate along the compactified dimension.  This is nothing
but a shift around half the circumference of the (dualized) thermal circle,
so that the states which are invariant under $\calT$ are those with 
even winding numbers.
This will then necessarily re-introduce states with odd winding numbers
in the twisted sectors, along with states having half-integer momentum numbers. 

Given these operators, the final step in our procedure is to orbifold the
circle-compactified theory in Eq.~(\ref{parfunct})
by the $\IZ_2$ product operator $\calT Q$.
What does this do to our partition function?
While $Q$ acts on the original non-thermal component $Z(\tau)$,
the operator $\calT$ acts on the thermal sum $Z_{\rm circ}(\tau,T)$.
Since states contributing 
to $Z_{\rm circ}$ with even (odd) values of $n$ are 
even (odd) under $\calT$, let us 
distinguish the specific values of $m$ and $n$ by 
introducing~\cite{Rohm} four new thermal functions $\calE_{0,1/2}$ and $\calO_{0,1/2}$
which are the same as the summation in
$Z_{\rm circ}$ in Eq.~(\ref{Zcircdef}) except for the following
restrictions on their summation variables:
\beqn
       \calE_0 &=& \lbrace  m\in\IZ,~n~{\rm even}\rbrace\nonumber\\
       \calO_0 &=& \lbrace  m\in\IZ,~n~{\rm odd}\rbrace\nonumber\\
       \calE_{1/2} &=& \lbrace  m\in\IZ+\half ,~n~{\rm even}\rbrace\nonumber\\
       \calO_{1/2} &=& \lbrace  m\in\IZ+\half ,~n~{\rm odd}\rbrace~.
\label{EOfunctions}
\eeqn
Note that $Z_{\rm circ}=\calE_0+\calO_0$.
Given this, our original (untwisted) thermal partition function
in Eq.~(\ref{parfunct}) 
can be rewritten as
\beq
         Z_{\rm therm,+}^+ ~=~ Z_+^+\,\left(\calE_0+\calO_0\right)~
\label{untwisted}
\eeq
where $Z_+^+(\tau)\equiv Z(\tau)$.
Therefore, in order to project onto the states invariant under $\calT Q$, we
add to Eq.~(\ref{untwisted})
the contributions from the projection sector
\beq 
        Z_{{\rm therm},+}^- ~=~  Z^-_+\,(\calE_0-\calO_0)
\eeq 
where $Z^-_+$ is the $Q$-projection sector for the 
non-thermal contribution $Z^+_+$.
In the usual fashion, modular invariance then requires us
to add the contribution from the twisted sector
\beq
        Z_{{\rm therm},-}^+ ~=~  Z^+_-\,(\calE_{1/2}+\calO_{1/2})~
\eeq
as well as its corresponding projection sector
\beq
        Z_{{\rm therm},-}^- ~=~  Z^-_-\,(\calE_{1/2}-\calO_{1/2})~.
\eeq
The net result of the orbifold, then, is a $(D-1)$-dimensional 
thermal string model with total partition function
\beqn
      Z_{\rm string}(\tau,T) &=&  {1\over 2}\, 
     \left( Z_{\rm therm,+}^+ + Z_{\rm therm,+}^- + 
            Z_{\rm therm,-}^+ + Z_{\rm therm,-}^-   \right)\nonumber\\ 
    &=& 
      {1\over 2}\, \biggl\lbrace
      \calE_0\,(Z^+_+ + Z^-_+) ~+~
      \calE_{1/2}\,(Z^+_- + Z^-_-) ~+~
      \calO_0\,(Z^+_+ - Z^-_+) ~+~ 
      \calO_{1/2}\,(Z^+_- - Z^-_-) \biggr\rbrace ~.
\label{mainresult}
\eeqn

It is straightforward to interpret the physics of this thermal model.
As $T\to \infty$, we find that
$\calE_{1/2}$ and $\calO_{1/2}$ each vanish while $\calE_0$ and $\calO_0$
become equal;
thus the partition function of our thermal model 
reduces to 
\beq
      Z_{\rm string}(\tau,T) ~\to~   Z^+_+ \equiv Z~~~~~~~{\rm as}~~T\to\infty~.
\eeq
In other words, we see that the original $D$-dimensional  model 
with which we started
can now be interpreted as the $T\to\infty$ limit of 
the $(D-1)$-dimensional thermal model we have constructed.
By contrast, as $T\to 0$, we find that $\calO_0$ and $\calO_{1/2}$ each vanish while 
$\calE_0$ and $\calE_{1/2}$ become equal.  Thus
\beq
      Z_{\rm string}(\tau,T) ~\to~   \half\left( Z^+_+ + Z_+^- + Z_-^+ + Z_-^-\right)  
                ~~~~~~~{\rm as}~~T\to 0~.
\eeq
However, this is nothing but the $Q$-orbifold of the original $D$-dimensional model with which
we began.  
Of course, since $Q$ is a $\IZ_2$ operator, we know that $Q^2={\bf 1}$.
We may therefore change our perspective and equivalently view our $D$-dimensional 
$T\to \infty$ model as the $Q$-orbifold of our 
$D$-dimensional $T\to 0$ model.
 
Thus, to summarize, we see that all finite-temperature string models 
must have partition functions of the modular-invariant form~\cite{Rohm,AtickWitten}
\beq
        Z_{\rm string}(\tau,T)  ~=~ 
           Z^{(1)}(\tau) ~ \calE_0(\tau,T) ~+~
           Z^{(2)}(\tau) ~ \calE_{1/2}(\tau,T) ~+~
           Z^{(3)}(\tau) ~ \calO_{0}(\tau,T) ~+~
           Z^{(4)}(\tau) ~ \calO_{1/2}(\tau,T) ~,
\label{EOmix}
\eeq
where $Z^{(i)}$ represent general, model-specific, non-thermal contributions to the total thermal
partition function $Z_{\rm string}$.
In the $T\to 0$ limit, 
we obtain a partition function of the form
\beq
          Z_{\rm model} ~=~ Z^{(1)} + Z^{(2)}~,
\label{origmodel}
\eeq
and thus we may interpret Eq.~(\ref{EOmix}) as describing the finite-temperature
extrapolation of the zero-temperature model described in Eq.~(\ref{origmodel}).
The fact that the orbifold
$Q$ contains a $(-1)^F$ factor guarantees that finite-temperature effects
will break whatever supersymmetry might have existed at zero temperature.
By contrast, the opposite  $T\to \infty$ limit yields
\beq
   \tilde Z_{\rm model} ~=~ Z^{(1)} + Z^{(3)}~,
\label{tildeZ}
\eeq
which corresponds to a different $D$-dimensional string model.
Thus, the thermal partition function in Eq.~(\ref{EOmix}) can be viewed
as mathematically {\it interpolating}\/ between one zero-temperature string model at $T=0$
[whose partition function is given in Eq.~(\ref{origmodel})]
and a {\it different}\/ zero-temperature string model as $T\to\infty$
[whose partition function is given in Eq.~(\ref{tildeZ})].
These two models are related directly in $D$ dimensions through the action
of the $\IZ_2$ orbifold operator $Q$.

This is a general result, so it bears repeating:
 {\it All $D$-dimensional thermal models are $(D-1)$-
dimensional interpolating models, with
the temperature $T$ serving as an 
interpolating parameter.
As $T\to 0$, we obtain a $D$-dimensional 
string model $M_1$;  this is identified as the zero-temperature string model whose 
thermal extrapolation we have constructed.
By contrast, as $T\to \infty$, we obtain a different $D$-dimensional
string model $M_2$ which must be a $\IZ_2$ orbifold of $M_1$.}

   A comment on semantics is in order here.  Strictly speaking, in the $T\to\infty$ 
   limit we obtain a $(D-1)$-dimensional degenerate (\ie, zero-radius) model $M_2$ which is actually only T-dual 
   to a $D$-dimensional model.  Thus, if $M_2$ is the $T\to\infty$ limit of our
   ($D-1$)-dimensional thermal interpolating model, then we should more correctly state that   
   our ($D-1$)-dimensional thermal model interpolates between the $D$-dimensional models 
   $M_1$ and $\tilde M_2$, where $\tilde M_2$ is the T-dual of $M_2$.  
   In some sense, this distinction is only a matter of semantics, having to
   do with the naming of the $T\to\infty$ endpoint of the interpolation;  moreover,
   for closed strings we should properly regard both $M_2$ and $\tilde M_2$ as being
   $D$-dimensional since they each have a continuous spectrum of states associated
   with the formerly compactified dimension. 
   For simplicity, therefore, we shall continue to refer to such an interpolating model as connecting
   $M_1$ and $M_2$ in the remainder of this paper.
   However, it is important to note that it is $M_2$ (and not $\tilde M_2$) which 
   must be the $Q$-orbifold of $M_1$.

Note that  for each specified $D$-dimensional zero-temperature string 
model $M_1$, there will in general exist many $(D-1)$-dimensional
string models which extrapolate away from it. 
This depends on the
choice of the second model $M_2$ to which one interpolates,
or equivalently on the choice of the $\IZ_2$ orbifold $Q$.
In other words, the requirement of modular invariance alone is not sufficient
to determine a unique interpolation, and is therefore not sufficient
to determine a unique thermal extrapolation.
The next step, then, is to determine a set of criteria for 
selecting the appropriate orbifold $Q$.

%=============================================================================
\section{What defines a self-consistent finite-temperature extrapolation?}
\setcounter{footnote}{0}

Let us now enumerate what we believe are the weakest possible conditions
that can be imposed in order to have a 
self-consistent finite-temperature
extrapolation of a given $D$-dimensional zero-temperature string model.
Throughout, our goal is to impose only the most conservative requirements
for self-consistency.
The conditions we shall impose are as follows:
\begin{itemize}
\item  First, the finite-temperature extrapolation should represent a
       valid, self-consistent $(D-1)$-dimensional string model in its own right.  
       In other words, it should satisfy all necessary worldsheet constraints
       such as conformal and superconformal invariance, self-consistent
       GSO projections, zero-temperature spin-statistics relations, {\it etc}\/.
\item  Second, this model should have an 
       identifiable radius modulus corresponding to a {\it bona-fide}\/ geometric 
        compactification circle.
\item  Finally, this compactification circle should be interpretable {\it thermally}\/
       in the field-theory limit.  This means that all states which survive in the 
       field-theory limit should satisfy field-theoretic {\it thermal}\/ spin-statistics
       relations.  
\end{itemize}
In practice, this last requirement means that all massless spacetime
bosons (fermions) with zero thermal windings in the theory 
should have thermal momentum excitations which are periodic (anti-periodic)
around the thermal circle.
Note, in particular, that we do {\it not}\/ make any demands on the
states with non-zero thermal winding modes;
such stringy states have no field-theoretic limits, and are beyond our usual
expectations.  Likewise, by restricting our attention to only the {\it massless}\/ states,
we are again focusing
on only the light states which can emerge in an appropriate
low-energy field-theoretic limit.
We stress that such {\it thermal}\/ spin-statistics relations should be contrasted
with the {\it zero-temperature}\/ spin-statistics relations
mentioned in the first of our requirements, which demand only 
that spacetime bosons (fermions) contribute with a positive (negative) 
sign to the overall partition function.

Given these conditions, we can now examine how well 
our construction in Sect.~I fares.
Clearly, our first condition
is automatically satisfied for {\it any}\/ orbifold $Q$, since the construction 
of our thermal model in Sect.~I proceeded by legitimate string-theoretic 
steps such as compactification and orbifolding.
As long as our original zero-temperature model is self-consistent and as long
as the orbifold $Q$ is a legitimate allowed orbifold for this model,
we are guaranteed that the resulting thermal model satisfies the first constraint.
Similarly, the second constraint is also satisfied, again by construction. 

However, the third constraint is more subtle.  At first glance, it might seem that
we have also satisfied our third constraint when we assumed that $Q$ contains 
a $(-1)^F$ factor and coupled it with the half-shift $\calT$ when constructing
our thermal orbifold.  However, there are two reasons  
why this may fail to be the case.
First, our orbifold $Q$ may generally contain other factors 
beyond $(-1)^F$ (such as gauge-group Wilson lines).  Thus,
in such cases, the thermal circle periodicities would be correlated not 
with spacetime fermion number alone, but with a combination of 
spacetime fermion number and Wilson-line
eigenvalues.  Of course, one might attempt to avoid this by taking $Q=(-1)^F$
directly, with no additional Wilson-line factors.  In cases when this can be done,
this issue will not arise, but we shall see shortly that this cannot always be done.
The second reason, however, is more general.  Recall that in our construction 
in Sect.~I, we began with a $D$-dimensional string model which turned out to
be the $T\to\infty$ limit of the thermal model we eventually constructed.
Indeed, it was only the $Q$-orbifold of this model which emerged in the 
zero-temperature limit.  Therefore, when we implemented the orbifold
factor $(-1)^F$ in our construction, 
this acted on the bosons/fermions of
the $T\to\infty$ model, but not necessarily those of the $T\to 0$ model.  
In other words, if a model interpolating from $M_1$ to $M_2$ is the proper
thermal extrapolation for the zero-temperature model $M_1$ (obeying proper
thermal spin-statistics relations for the bosons and fermions of $M_1$),
there is no guarantee that the T-dual model, which interpolates from 
$M_2$ to $M_1$, will be the appropriate thermal extrapolation for the model $M_2$.
Thus, the construction we outlined in Sect.~I --- although completely general --- is not
by itself capable of guaranteeing that  we have successfully maintained proper thermal 
spin-statistics relations when the final $(D-1)$-dimensional interpolation
is constructed.
Note that this remains true even in cases when $Q=(-1)^F$.

It is therefore this third requirement involving proper thermal spin-statistics
relations which can be used to 
select the proper orbifold $Q$, and with it the correct
thermal extrapolation for a given string model.
We shall give two explicit examples of this procedure below.

%=============================================================================
\section{Relation to the standard approach}
\setcounter{footnote}{0}

Let us now compare how our discussion thus far relates to the standard
prescription in the literature~\cite{AtickWitten} for 
constructing the thermal extrapolation of a given zero-temperature
string model.  
We shall explicitly work out two examples in ten dimensions:
the supersymmetric Type~IIA/IIB superstrings, and the
supersymmetric $SO(32)$ heterotic string.
As we shall show, the standard prescription 
agrees completely with the
our construction
in the case of the Type~II superstrings,
but differs in the case of the $SO(32)$ heterotic string.
In the latter case, we shall explicitly describe exactly where
we believe the problem lies with the usual prescription.

\subsection{The Type II strings}

Let us begin by considering the case of the 
ten-dimensional Type~II superstrings.
For concreteness, we shall focus on the (chiral) Type~IIB
string;  the case of the (non-chiral) Type~IIA string 
proceeds in exactly the same manner.
The Type~IIB string at zero temperature has the partition function
\beqn
      Z_{\rm IIB} &=& Z_{\rm boson}^{(8)} ~
           (\chibar_V-\chibar_S)\,(\chi_V-\chi_S) ~
\label{IIB}
\eeqn
where the contribution from the worldsheet bosons is given in terms of the Dedekind $\eta$-function
as
\beq
         Z^{(n)}_{\rm boson} ~\equiv ~ {\tau_2}^{-n/2}\, (\overline{\eta}\eta)^{-n}~,
\label{bosons}
\eeq
and where the contributions from the left--moving (right-moving) worldsheet fermions are 
written in terms of the unbarred (barred) characters $\chi_i$ ($\chibar_i$) 
of the transverse $SO(8)$ Lorentz
group.  In general, the subscripts $I$, $V$, $S$, and $C$ refer 
to the identify, vector, spinor, and conjugate
spinor representations of any $SO(2n)$ affine Lie group;
these representations have conformal dimensions $\lbrace h_I,h_V,h_S,h_C\rbrace=
\lbrace 0,1/2,n/8,n/8\rbrace$, and have corresponding characters
which can be expressed in terms of Jacobi $\vartheta$-functions as
\beqn
 \chi_I &=&  \half\,(\tthree^n + \tfour^n)/\eta^n ~=~ q^{h_I-c/24} \,(1 + n(2n-1)\,q + ...)\nonumber\\
 \chi_V &=&  \half\,(\tthree^n - \tfour^n)/\eta^n ~=~ q^{h_V-c/24} \,(2n + ...)\nonumber\\
 \chi_S &=&  \half\,(\ttwo^n + {\vartheta_1}^n)/\eta^n ~=~ q^{h_S-c/24} \,(2^{n-1} + ...)\nonumber\\
 \chi_C &=&  \half\,(\ttwo^n - {\vartheta_1}^n)/\eta^n ~=~ q^{h_C-c/24} \,(2^{n-1} + ...)~
\label{chis}
\eeqn
where the central charge is $c=n$ at affine level one.
For the ten-dimensional transverse Lorentz group $SO(8)$,
the distinction between $S$ and $C$ is equivalent to relative
spacetime chirality.
Note that the $SO(8)$ transverse Lorentz group
has a triality symmetry under which
the vector and spinor representations are
indistinguishable.  Thus $\chi_V=\chi_S$ and $\overline{\chi}_V=\overline{\chi}_S$, 
resulting in a (vanishing) supersymmetric partition function in Eq.~(\ref{IIB}).
The presence of two such factors in Eq.~(\ref{IIB}) reflects the $\calN=2$ supersymmetry
of this model at zero temperature.

Let us now consider the extrapolation of this theory to finite temperature.
If we apply the the standard prescription in Ref.~\cite{AtickWitten},
we obtain a nine-dimensional extrapolation with partition function
\beqn
    Z_{\rm string}(\tau,T) ~=~  Z^{(7)}_{\rm boson} \,\times \,\bigl\lbrace ~
    \phantom{+}&\calE_0 &  \lbrack
     \chibar_V\chi_V + \chibar_S\chi_S
        \rbrack\nonumber\\
   -&\calE_{1/2}  & \lbrack
      \chibar_V\chi_S + \chibar_S\chi_V
    \rbrack\nonumber\\
   +&\calO_0 & \lbrack
    \chibar_I\chi_I + \chibar_C\chi_C
     \rbrack\nonumber\\
   -&\calO_{1/2} & \lbrack
     \chibar_I\chi_C + \chibar_C\chi_I
  \rbrack ~~\bigr\rbrace~ .
\label{TypeIIinterp}
\eeqn
It is easy to interpret this partition function in the language
of our construction in Sect.~I, and in the process verify that this result
is self-consistent.  First, we see that the $T\to 0$ limit of this
expression reproduces the Type~IIB partition function in Eq.~(\ref{IIB}),
while the $T\to \infty$ limit of this expression yields the partition function
of the non-supersymmetric ten-dimensional Type~0B superstring:
\beqn
      Z_{\rm 0B} &=& Z_{\rm boson}^{(8)} ~
         (\chibar_I\chi_I + \chibar_V\chi_V + \chibar_S\chi_S + \chibar_C\chi_C) ~.
\eeqn
Thus, the nine-dimensional thermal model in Eq.~(\ref{TypeIIinterp}) interpolates
between the Type~IIB string and the Type~0B string,
thereby breaking supersymmetry for all $T>0$. 
(The extra factor of $Z_{\rm boson}^{(1)}$ required in these limits emerges as the 
limit of the thermal $\calE/\calO$ functions.)  
Note, moreover, that the Type~0B string is nothing but a $\IZ_2$ orbifold 
of the Type~IIB string, where the orbifold action is simply $Q=(-1)^F$ where
$F$ denotes the total spacetime fermion number.  This is indeed a legitimate orbifold
action for the Type~IIB string, and is in fact the only such orbifold that would have
been possible for the Type~IIB theory. 
Finally, we note that this interpolation also satisfies proper thermal spin-statistics
relations in the field-theoretic limit.  Specifically, massless spacetime bosons 
(such as those arising within the sectors $\chibar_V \chi_V$ or $\chibar_S\chi_S$) 
have integer modings around the thermal circle (\ie, they are multiplied by $\calE_0$),
while massless spacetime fermions (such as within $\chibar_V\chi_S$ or
$\chibar_S\chi_V$) have half-integer modings around the thermal circle (\ie, they
are multiplied by $\calE_{1/2}$).  As discussed in Sect.~II, 
we do not impose any requirements on massive states or states with non-zero windings
(such as those corresponding to the $\calO_0$ or $\calO_{1/2}$ sectors), but we observe that the usual
thermal spin-statistics relations actually hold in those sectors as well.

We conclude, then, that the standard prescription agrees completely with our
construction for thermal string models in the case of the Type~IIB string.
The case of the Type~IIA string is almost exactly the same;  we can simply replace
$\chi_S\leftrightarrow \chi_C$ for the left-moving characters throughout the above
expressions.  The Type~IIA thermal extrapolation is therefore one which interpolates
between the Type~IIA string at $T=0$ and the Type~0A string as $T\to\infty$.

Note that in each case, these extrapolations are tachyon-free up to a certain
critical temperature.  As we shall discuss below, their failure to remain tachyon-free beyond
this temperature is the signal for the usual Type~II
Hagedorn transition.

\subsection{The heterotic theory}

Let us now turn to the case of the ten-dimensional $SO(32)$ heterotic string.
It is here that we shall find an important difference relative to the usual
thermal prescription.
The ten-dimensional $SO(32)$ heterotic string
has the zero-temperature partition function
\beq
         Z_{SO(32)} ~=~ Z^{(8)}_{\rm boson}~
         (\overline{\chi}_V-\overline{\chi}_S)
         \,(\chi_I^2 + \chi_V^2 + \chi_S^2 + \chi_C^2)~.
\label{SO32partfunct}
\eeq
As with the Type~II string,
the contribution from the worldsheet bosons is given in Eq.~(\ref{bosons})
and the contributions from the right-moving worldsheet fermions are written in terms of
the barred characters $\chibar_i$ of the transverse $SO(8)$ Lorentz
group.  The major new notational difference in the heterotic case is that 
the contributions from the left-moving (internal) worldsheet fermions  
are now written as products of the unbarred characters $\chi_i$ of an internal $SO(16)$ 
gauge group.  Thus, for notational convenience, we are essentially decomposing 
our $SO(32)$ characters into the characters of the subgroup $SO(16)\times SO(16)$.

Let us now consider the extrapolation of this theory to finite temperature.
If we were to apply the standard prescriptions,
we would obtain a nine-dimensional extrapolation with
partition function
\beqn
    Z ~=~  Z^{(7)}_{\rm boson} \,\times \,\bigl\lbrace ~
         \phantom{+}&\calE_0 &  \chibar_V ~(\chi_I^2 + \chi_V^2 + \chi_S^2 + \chi_C^2) \nonumber\\
         -&\calE_{1/2} &  \chibar_S ~(\chi_I^2 + \chi_V^2 + \chi_S^2 + \chi_C^2)\nonumber\\
         -&\calO_{0} &  \chibar_C ~(\chi_I^2 + \chi_V^2 + \chi_S^2 + \chi_C^2)\nonumber\\
         +&\calO_{1/2} &  \chibar_I ~(\chi_I^2 + \chi_V^2 + \chi_S^2 + \chi_C^2)~\bigr\rbrace~.
\label{fake}
\eeqn
This is indeed modular invariant, and 
incorporates proper thermal spin-statistics relations
for states with zero thermal windings (\ie, states multiplying 
the $\calE_0$ and $\calE_{1/2}$ thermal functions).
Despite these successes, it is easy to demonstrate that Eq.~(\ref{fake}) 
cannot represent a self-consistent
thermal extrapolation according to the construction procedure we laid out in 
Sect.~I --- \ie, that this cannot be the partition function of a 
 {\it bona-fide}\/ self-consistent
nine-dimensional string model.

Before giving a definitive argument to this effect, let us begin by noting 
that there are certain immediate clues that all is not well.
First, we observe
that Eq.~(\ref{fake}) would appear to represent a non-supersymmetric
interpolation between
two {\it supersymmetric}\/ limits, one at $T=0$ and the other at $T\to \infty$,
both of which represent the same $SO(32)$ heterotic string model but with opposite
spacetime chiralities! 
Indeed, while the $T\to 0$ limit of Eq.~(\ref{fake}) reproduces
Eq.~(\ref{SO32partfunct}), the $T\to \infty$ limit yields
\beq
         Z'_{SO(32)} ~=~ Z^{(8)}_{\rm boson}~
         (\overline{\chi}_V-\overline{\chi}_C)
         \,(\chi_I^2 + \chi_V^2 + \chi_S^2 + \chi_C^2)~.
\label{SO32partfunctflipped}
\eeq
This is precisely the same model with which we started at zero temperature,
only now involving spacetime spinors of opposite chirality.
This is quite unlike the case of the Type~II string, where the $T\to\infty$
endpoint model was the non-supersymmetric Type~0A or 0B theory. 

Second, we observe that in Eq.~(\ref{fake}), 
the string worldsheet CFT ground state character $\chibar_I \chi_I^2$
appears within the sector multiplied by $\calO_{1/2}$.  
To see why this is a problem, let us first observe that
it follows from modular invariance that this is the only
place such a term could possibly have appeared:
since the ground state of the heterotic string is not level-matched,
having worldsheet energies $(H_R,H_L)=(-1/2,-1)$,
invariance under $\tau\to \tau+1$
forces such a term to be multiplied
by the function $\calO_{1/2}$, which also fails to be level-matched.  
In other words, modular invariance requires a term such as $\chibar_I\chi_I^2$, 
if it exists,
to appear multiplied by $\calO_{1/2}$ rather than by any of the other thermal
functions.
However, this is a problem because the $\calO_{1/2}$ sector must be interpreted as 
completely {\it twisted}\/ with respect to the $\IZ_2$ thermal orbifold.   
Indeed, no matter whether we run our interpolations from $T\to 0$ to $T\to\infty$
or backwards from $T\to \infty$ to $T\to 0$, the contributions multiplying
$\calO_{1/2}$ can only correspond to twisted sectors.
However, we do not expect to see the ground state of a self-consistent 
conformal field theory emerging from a twisted sector.
Equivalently stated, we expect a term of the form $\chibar_I\chi_I^2$
to appear multiplied by $\calE_0$, $\calE_{1/2}$, or $\calO_0$, but never $\calO_{1/2}$.
Thus, combining these observations,  
we see that a self-consistent heterotic thermal extrapolation
should not have a term of the form $\chibar_I \chi_I^2$ appearing {\it anywhere}\/
in its partition function.  Yet, this term appears within Eq.~(\ref{fake}).

In order to diagnose the source of the problem, 
let us return to our original observation that the nine-dimensional ``model'' in
Eq.~(\ref{fake}) seems to
interpolate between to supersymmetric endpoints:  the $SO(32)$ heterotic string
at zero temperature, and the chirality-flipped $SO(32)$ heterotic string
at infinite temperature.  However, according to our discussion in Sect.~I,
this can represent a consistent nine-dimensional interpolation
only if the chirality-flipped $SO(32)$ model can be viewed as a $\IZ_2$ orbifold
of the unflipped $SO(32)$ model.  
We shall now demonstrate that there is no such $\IZ_2$ orbifold which can accomplish
this transformation.

To see this, let us consider the worldsheet sector giving rise to the gravitino
of the original supersymmetric $SO(32)$ model.
Recall that in the heterotic string, the gravitino state is realized 
in the Ramond sector as the spin-3/2 component within the tensor product 
\beq
        \hbox{gravitino:}~~~~~~~~~~~~
             \tilde g^{\alpha \nu} ~\subset ~
         \lbrace \tilde b_{0}\rbrace^\alpha  |0\rangle_R ~\otimes~ \alpha_{-1}^\nu |0\rangle_L~.
\label{gravitino}
\eeq
Here $\alpha^\nu_{-1}$ denotes the lowest excitation
of the left-moving worldsheet coordinate boson $X^\nu$,
with its Lorentz vector index $\nu$,
while 
$\lbrace \tilde b_{0}\rbrace^\alpha$ schematically indicates the Ramond zero-mode
combinations which collectively give rise to the spacetime Lorentz spinor index $\alpha$,
as required for the spin-3/2 gravitino state.
Note that in order for such a state to be massless and level-matched,
it must emerge from a sector in which
the left-moving (conformal) side of the heterotic
string is in the completely Neveu-Schwarz ground state, while the right-moving (superconformal)  
side of the heterotic string is in the completely Ramond ground state.
Note that in ten dimensions, 
this is the {\it only}\/ sector which can ever give rise to spacetime gravitinos,
and as such this sector is unique.

So what must our desired orbifold do?
In general, orbifolds project certain states out of the spectrum from untwisted
sectors, but then introduce new twisted sectors from which additional states
may emerge.  In our case, the desired orbifold must 
project out the gravitino that previously emerged from 
the gravitino sector.  This is because the $Q$-orbifold must break spacetime supersymmetry
(or alternatively, because the gravitino has the wrong chirality).
However, our orbifolding 
procedure must also somehow restore an opposite-chirality gravitino from a twisted sector.
This is necessary so that the net result of the orbifolding procedure can be the 
chirality-flipped supersymmetric $SO(32)$ model.

Ordinarily, there are many instances in which a given state might be projected out of
the spectrum from an untwisted sector only to re-emerge from a twisted sector.
However, as we have noted above, 
the gravitino sector is {\it unique}\/
within the context of ten-dimensional heterotic strings  ---
this is the only sector which can provide gravitinos of either chirality.
It has a unique worldsheet construction,
and thus no other sector can re-introduce the needed opposite-chirality   
gravitino once the original gravitino has been projected out of the 
original untwisted sector.
We conclude, then, that there is no self-consistent $\IZ_2$ orbifold $Q$ which can possibly
transform the ten-dimensional supersymmetric $SO(32)$ heterotic string into a
chirality-flipped version of itself.  It then follows from the construction presented
in Sect.~I that there is no self-consistent nine-dimensional interpolation between 
these two ten-dimensional endpoints.

Note that we are not saying that an orbifold cannot project out certain states
from an untwisted sector, only to have them re-emerge (even with chirality flips)
from a twisted sector.  This indeed happens quite often.  Rather, we 
are claiming that for the gravitino in ten dimensions, the actual
 {\it worldsheet sector}\/ from which such a state can arise is unique, and thus such
a sector cannot be both untwisted and twisted with respect to the same orbifold.  
In other words, each sector can contribute only once in a given string model.

    Note that in this discussion, we are identifying our endpoint models
    $M_1$ and $M_2$ as the supersymmetric $SO(32)$ heterotic string models with 
    opposite chiralities, in accordance with the results of Sect.~I.~ 
    Since $M_1=\tilde M_2$ in this case (where $\tilde M_2$ is the T-dual of $M_2$),
    one might instead try to form the desired thermal model
    by interpolating between $M_1$ and itself, \ie, between $M_1$ and $\tilde M_2$,
    thereby avoiding the chirality flip.
    However, the chirality flip is not the real issue here.
    Indeed, the original $SO(32)$ heterotic gravitino 
    must always be projected out
    of the spectrum by the $Q$-orbifold;  otherwise, supersymmetry would not be broken
    by thermal effects in such an interpolation.  A new gravitino cannot then be 
    re-introduced from a twisted sector, regardless of its chirality.  

 {\it We conclude, then, that  Eq.~(\ref{fake}), although modular invariant, fails to
represent a self-consistent thermal extrapolation of the ten-dimensional $SO(32)$
heterotic string.  In particular, it cannot correspond to a self-consistent nine-dimensional
interpolating string model at the worldsheet level.}\/  Identical arguments 
apply as well to the $E_8\times E_8$ heterotic string.

%================================================================================
\section{Correct finite-temperature extrapolation for the $SO(32)$ heterotic string}

In order to derive the correct finite-temperature extrapolation for the $SO(32)$
heterotic string, we follow our procedure in Sect.~I.~  Specifically, we   
must choose an appropriate orbifold $Q$ of the $SO(32)$ string, and then develop
the corresponding nine-dimensional interpolating model.

What are the possible self-consistent $\IZ_2$ orbifolds of the  $SO(32)$ heterotic
string?  Clearly, this question boils down to the question of identifying 
possible ten-dimensional $T\to \infty$ endpoints for our corresponding nine-dimensional
interpolation.  Fortunately, all heterotic strings in ten dimensions have  
been classified~\cite{KLTclassification}.  In addition to the supersymmetric
$SO(32)$ and $E_8\times E_8$ heterotic strings, there are seven additional
non-supersymmetric strings.  These are the tachyon-free $SO(16)\times SO(16)$
string model as well as six tachyonic string models with 
gauge groups $SO(32)$, $SO(8)\times SO(24)$, $U(16)$,
$SO(16)\times E_8$,
$(E_7)^2 \times SU(2)^2$,  and
$E_8$.
The tachyons in the latter six models all have
worldsheet energies $(H_R,H_L)=(-1/2,-1/2)$.
However,
not all of these models can be realized as 
$\IZ_2$ orbifolds of the supersymmetric $SO(32)$ model, and
even in the remaining cases,
not all of the resulting interpolating models will have a radius of compactification
that can be interpreted thermally in the field-theory limit, as required by our third
condition in Sect.~II.

Fortunately, one interpolation meets all of our requirements.  Perhaps
not surprisingly, this is the
interpolation between the supersymmetric $SO(32)$ string and the non-supersymmetric
$SO(32)$ string.  Note that the non-supersymmetric $SO(32)$ heterotic string has
partition function
\beq
       Z ~=~ Z^{(8)}_{\rm boson}~\times \biggl\lbrace 
    \chibar_I  \,(\chi_I\chi_V+\chi_V\chi_I) ~+~ \chibar_V \,(\chi_I^2 + \chi_V^2) 
       -~\chibar_S\, (\chi_S^2 + \chi_C^2) ~- ~\chibar_C \, ( \chi_S\chi_C+\chi_C\chi_S)
          ~\biggr\rbrace~.
\label{nonsusyso32}
\eeq
Following the procedure outlined in Sect.~I,
we then find that 
the corresponding nine-dimensional interpolating model
has the partition function~\cite{julie}
\beqn
    Z_{\rm string}(\tau,T) ~=~  Z^{(7)}_{\rm boson} \,\times \,\bigl\lbrace ~
    \phantom{+}&\calE_0 &
     \lbrack \chibar_V \,(\chi_I^2 + \chi_V^2)  ~-~  \chibar_S \,(\chi_S^2 + \chi_C^2) \rbrack\nonumber\\
   +&\calE_{1/2}  &
       \lbrack \chibar_V \,(\chi_S^2+\chi_C^2)  ~-~  \chibar_S \,(\chi_I^2+\chi_V^2) \rbrack\nonumber\\
   +&\calO_0 &
      \lbrack \chibar_I \,(\chi_I\chi_V+\chi_V\chi_I) ~-~ \chibar_C \,(\chi_S\chi_C+\chi_C\chi_S) \rbrack\nonumber\\
   +&\calO_{1/2} &
       \lbrack \chibar_I \,(\chi_S\chi_C+\chi_C\chi_S) ~-~ \chibar_C \,(\chi_I\chi_V+\chi_V\chi_I) \rbrack ~~\bigr\rbrace~.
\label{interp1}
\eeqn
Note, in particular, that this correctly reproduces 
Eq.~(\ref{SO32partfunct}) in the $T\to 0$ limit
as well as Eq.~(\ref{nonsusyso32}) in the $T\to\infty$ limit.
Moreover, it satisfies thermal spin-statistics relations for the
massless states with zero string windings:  all massless states multiplying
$\calE_0$ are spacetime bosons, while all massless states multiplying
$\calE_{1/2}$ are spacetime fermions.
(Note in this context there are no massless states
which contribute to terms of the 
forms $\chibar_{S,V} \chi_{S,C}^2$,
since $\chi_{S,C}^2$ has conformal dimension
$h=2$.)

We stress that this is the unique thermal extrapolation which satisfies the conditions
we put forth in Sect.~II.  ~Indeed, only this extrapolation corresponds to a self-consistent
nine-dimensional interpolating model with an identifiable thermal radius of compactification
with proper thermal spin-statistics relations in the field-theory limit. 
However, there are some unique features involved in such an interpolation.  
While it was perhaps already expected from Ref.~\cite{AtickWitten} that states with
non-trivial thermal winding modes
might behave in a counter-intuitive fashion, violating finite-temperature
spin-statistics relations in the $\calO_0$ and $\calO_{1/2}$ sectors,
we now see that our interpolations necessarily have apparent 
Planck-scale thermal spin-statistics violations 
even for the states with {\it zero}\/ windings, \ie, states with 
conformal dimensions $h>1$ which appear in the $\calE_0$
and $\calE_{1/2}$ sectors.
Planck-scale violations of this sort appear to be unavoidable, even for zero-winding states,
and (as we shall argue below) 
are required by modular invariance in the context of self-consistent
interpolating models.
It would be interesting to understand the thermal implications of these states as far as
Planck-scale physics is concerned.

Note, however, that although these Planck-scale
states appear to violate {\it thermal}\/ spin-statistics
relations, they still obey zero-temperature spin-statistics relations, as required.
In other words, all spacetime bosons contribute positively to the partition function,
while all spacetime fermions contribute negatively, with minus signs in front of
their corresponding characters.

%=============================================================================
\section{Implications for the Hagedorn transition}
\setcounter{footnote}{0}

Let us now discuss the implications of these results for the Hagedorn 
transition~\cite{Hagedorn,Polbook,AtickWitten}.
Our focus here will be on the tachyons and temperature associated with
the Hagedorn transition;  for a more complete set of references concerning
the history and possible interpretations and implications 
of the Hagedorn transition, see Ref.~\cite{DL}.

%=================================================================================
\subsection{The Hagedorn transition:  UV versus IR}

We begin with several preliminary remarks concerning the UV/IR nature of the 
Hagedorn transition.
In general, once we have determined the correct finite-temperature partition function
$Z_{\rm string}(\tau,T)$ for a given zero-temperature string model,
the one-loop thermal vacuum amplitude $\calV(T)$ (the analogue of the 
logarithm of the statistical-mechanical partition function) is given   
by a modular integral of the form~\cite{Pol86}
\beq
    \calV(T) ~\equiv~ -\half \,{\cal M}^{D-1}\, \int_{\cal F}
              {d^2 \tau\over ({\rm Im} \,\tau)^2}
             \,Z_{\rm string}(\tau,T)
\label{Vdef}
\eeq
where ${\cal M}\equiv M_{\rm string}/(2\pi)$ is the reduced string scale;
$D$ is the number of non-compact spacetime dimensions;
and where
\beq
   {\cal F}~\equiv ~\lbrace \tau:  ~|{\rm Re}\,\tau|\leq \half,
 {\rm Im}\,\tau>0, |\tau|\geq 1\rbrace
\label{Fdef}
\eeq
is the fundamental domain
of the modular group.  We shall 
often abbreviate $\tau_1\equiv {\rm Re}\,\tau$
and $\tau_2\equiv {\rm Im}\,\tau$.
Given this definition for $\calV$, the free energy $F$, internal
energy $U$, entropy $S$, and specific heat $c_V$ then follow
from the standard thermodynamic definitions $F \equiv T \calV$,
         $U \equiv - T^2 {d\calV /dT}$,
         $S \equiv  -{dF/dT}$,
and         $c_V \equiv {dU/ dT}$.

In string theory, 
the Hagedorn transition is usually associated with a divergence or other discontinuity
in the vacuum amplitude $\calV(T)$ as a function of temperature. 
There are two ways in which such a divergence may arise.
First, of course, there may be an ultraviolet divergence due to the well-known exponential
rise in the degeneracy of string states.  However, such an ultraviolet divergence
would normally be associated with the $\tau_2\to 0$ region of integration in Eq.~(\ref{Vdef}),
and we see from Eq.~(\ref{Fdef}) that this region of integration has been eliminated
from the integral as a result of modular invariance --- \ie, as the result of the
truncation of the region of integration to the modular group fundamental domain
in Eq.~(\ref{Fdef}).  Thus, strictly speaking, there
can be no UV divergence contributing to $\calV(T)$.  On the other hand, there may be
purely infrared divergences coming from on-shell physical tachyons within $Z_{\rm string}(\tau,T)$;
such states would lead to divergences in the infrared region $\tau_2\to\infty$.
Thus, a study of the Hagedorn transition in string theory essentially reduces 
to a study of the possible tachyonic structure of $Z_{\rm string}(\tau,T)$ as a function
of temperature.

Before proceeding further, we caution that we reach this conclusion only because
we have chosen to work in the so-called $\calF$-representation for $\calV(T)$ given
in Eq.~(\ref{Vdef}).
By contrast,
utilizing Poisson resummations and modular transformations~\cite{McClainRoth},
we can always rewrite
$\calV(T)$ as the integration of a different integrand $Z'_{\rm string}(\tau,T)$
over the strip
\beq
   {\cal S}~\equiv ~\lbrace \tau:  ~|{\rm Re}\,\tau|\leq \half, {\rm Im}\,\tau>0 \rbrace~.
\label{Sdef}
\eeq
In such an $\calS$-representation, the former IR divergence as $\tau_2\to\infty$ 
is transformed into a UV divergence
as $\tau_2\to 0$.  This formulation thus has the advantage of relating the Hagedorn
transformation directly to a UV phenomenon such as the exponential rise in the degeneracy of states.
However, both formulations are mathematically equivalent;  indeed, modular invariance
provides a tight relation between the tachyonic structure of a given partition function
and the rate of exponential growth in its asymptotic degeneracy of states~\cite{HR,Kani,missusy}.
In the following, therefore, we shall utilize the $\calF$-representation for 
$\calV(T)$ and focus on only the tachyonic structure of $Z_{\rm string}(\tau,T)$,
but we shall comment on the connection to the asymptotic degeneracy of states
in Sect.~V$\,$C.

%=================================================================================
\subsection{A new Hagedorn temperature for heterotic strings}

So what then are the potential tachyonic states within $Z_{\rm string}(\tau,T)$,
and at what temperature $T_H$ do they first arise?  In other words, at what critical
temperature $T_H$ do new massless states emerge within $Z_{\rm string}$, on their way to
becoming tachyonic? 
Note that we are focusing on {\it thermally massless}\/ states, \ie, states 
whose masses
depend on temperature, states 
for which masslessness is achieved at a critical temperature
$T_H$ as the result of a balance  between a
tachyonic non-thermal mass
and an additional positive non-zero thermal mass contribution.
It is sufficient to focus on such massless states
since their emergence is the signal of the long-range order normally associated
with a phase transition.
These are the states which then presumably become  
tachyonic beyond $T_H$, leading to the instabilities normally associated with a
phase transition.

Given our results for $Z_{\rm string}(\tau,T)$ in Sects.~III and IV, 
it is straightforward to obtain the corresponding
Hagedorn temperatures.
Let us begin by considering the case of the Type~II strings, for which
the appropriate thermal function is given in Eq.~(\ref{TypeIIinterp}). 
Recalling the conformal dimensions associated with the different characters
in Eq.~(\ref{TypeIIinterp}), we see that 
the only potentially tachyonic contributions in this expression arise from 
the term $\chibar_I\chi_I \calO_0$.  Thus, only this sector has the potential
to give rise to thermally massless level-matched states.  
Solving the conditions for masslessness, we find that 
the $(m,n)=(0,\pm 1)$ thermal excitations of the  
$(H_R,H_L)=(-1/2,-1/2)$ tachyons within $\chibar_I\chi_I$
will become thermally massless at the temperature $T_H=\calM/\sqrt{2}$.
These thermal states are massive for $T<T_H$, and tachyonic for $T>T_H$.
We thus identify $T_H=\calM/\sqrt{2}$ as the Hagedorn temperature for
Type~II strings.
Note that this analysis is in complete agreement with 
the standard derivations~\cite{Polbook,AtickWitten} of the Hagedorn temperature for Type~II strings.

However, the main difference occurs in the case of the heterotic string.
Performing exactly the same analysis for the thermal partition function
given in Eq.~(\ref{interp1}), we find that only the term 
$\chibar_I (\chi_I\chi_V+\chi_V\chi_I)\calO_0$  
is capable of
giving rise to thermally massless level-matched states.
Indeed, the $SO(16)\times SO(16)$ character 
$(\chi_I\chi_V+\chi_V\chi_I)$ gives rise to the 
32 on-shell $(H_R,H_L)=(-1/2,-1/2)$ tachyons of  
non-supersymmetric $SO(32)$ string which serves as the $T\to\infty$ endpoint
of the interpolation, and we find that the $(m,n)=(0,\pm 1)$ thermal excitations
of these states are massless at $T_H=\calM/\sqrt{2}$, massive below this temperature,
and tachyonic above it.
This is exactly the same as for the Type~II string, and there are no other
tachyonic sectors within Eq.~(\ref{interp1}) which could give rise to other
phase transitions at lower temperatures.
We thus conclude that the ten-dimensional supersymmetric $SO(32)$ heterotic
string has a Hagedorn temperature given by $T_H=\calM/\sqrt{2}$, which is exactly
the same as the Hagedorn temperature for the Type~II string.
A similar analysis with the same result also applies for the $E_8\times E_8$ heterotic
string as well as the non-supersymmetric tachyon-free $SO(16)\times SO(16)$ string.
We thus find that
\beq
           T_H ~=~ {\calM\over \sqrt{2}} ~=~ {M_{\rm string}\over 2\sqrt{2}\pi}~~~~~~~
      \hbox{for {\it all}\/ tachyon-free closed strings in $D=10$}~, 
\eeq 
both Type~II and heterotic!
In other words, by carefully constructing self-consistent interpolating models
with their required GSO projections, we have uncovered a universal Hagedorn temperature 
for all closed tachyon-free strings in ten dimensions.

This is clearly a major difference relative to our usual expectations.
Indeed, if we had performed the same analysis using the (inconsistent) 
expression in Eq.~(\ref{fake}), 
we would have found 
that the $(H_R,H_L)=(-1/2,-1)$ off-shell tachyon  
within the sector $\chibar_I \chi_I^2\calO_{1/2}$
contains thermal excitations $(m,n)=\pm(1/2,1)$
which become thermally massless at the expected (heterotic) Hagedorn temperature   
$T_H= 2\calM/(2+\sqrt{2}) = (2-\sqrt{2})\calM$.
However, as we discussed in Sect.~IV, this state is actually GSO-projected
out of the spectrum when we construct the proper thermal interpolating model
in Eq.~(\ref{interp1}).  We thus find that the Hagedorn temperature for the
heterotic string is altered.

It is not surprising, perhaps, that the Type~II and heterotic strings share a common
Hagedorn temperature once the correct thermal extrapolations are taken into account.
In the case of the Type~II string, the ground state is a tachyon with worldsheet
energies $(H_R,H_L)=(-1/2,-1/2)$.   This state is level-matched, and survives into
the corresponding thermal extrapolating function in Eq.~(\ref{TypeIIinterp}).
In the case of the heterotic string, 
by contrast,
the ground state has vacuum energies $(H_R,H_L)=(-1/2,-1)$.  Although this
would naively appear to change the associated Hagedorn temperature,
the fact that this state is not level-matched, together with modular invariance,
ends up forcing this state to appear within the thermally twisted sector $\calO_{1/2}$ 
where its appearance would be inconsistent.  Thus, all thermal 
contributions from this state are projected out, and only the ``next-deepest''
tachyon, again with $(H_R,H_L)=(-1/2,-1/2)$, survives to affect the resulting
thermodynamics.  Since this surviving heterotic tachyon has
exactly the same worldsheet energies as the 
Type~II ground state, the heterotic and Type~II theories have
exactly the same Hagedorn temperatures.

%=================================================================================
\subsection{Reconciling the new Hagedorn temperature with the asymptotic
    degeneracy of states}

The above arguments are clearly based on an IR analysis of the tachyonic
structure of our thermal interpolating models.
One may therefore wonder how it is possible to find a Hagedorn temperature 
$T_H=\calM/\sqrt{2}$ for
a heterotic string such as the $SO(32)$ string, given that  
its zero-temperature bosonic and fermionic densities of states 
nevertheless continue to exhibit an exponential rate of growth normally 
associated with the usual heterotic Hagedorn temperature $T_H=(2-\sqrt{2})\calM$.
This is a very important question which we shall now address.

We shall develop our response in several layers.
First, let us recall the 
usual direct connection 
between the asymptotic degeneracy of states and the 
corresponding Hagedorn temperature~\cite{Hagedorn}:
if $g_M$ denotes the number of states with mass $M$, then the thermal partition function is
given by $Z(T)=\sum  g_M e^{-M/T}$.  However, if $g_M \sim e^{\alpha M}$ as $M\to\infty$, then
$Z(T)$ diverges for $T\geq T_H\equiv 1/\alpha$.
This appears to provide a firm link between
the Hagedorn temperature and the asymptotic degeneracy of states.

Of course, one might argue that this kind of partition
function $Z(T)=\sum  g_M e^{-M/T}$ is not a proper string-theoretic
partition function;  it assumes that the string is nothing but a
collection of the states to which its excitations give rise.
Indeed, we must perform a proper string-theoretic vacuum-amplitude calculation
as outlined in Eq.~(\ref{Vdef}), using a string partition function $Z_{\rm string}(\tau,T)$ which
depends not only on the temperature $T$ but also a torus parameter $\tau$.
We must then integrate over $\tau$ over a restricted fundamental domain.

However, the same basic argument
connecting the asymptotic degeneracy of states
with the Hagedorn transition
continues to apply, even for the proper string-theory calculation.
After all, we may 
easily transform to the $\calS$-representation for $\calV(T)$, as discussed
in Sect.~V$\,$A;  
in this representation, the connection between the asymptotic
rise in the degeneracy of states and the UV divergence as $\tau_2\to 0$ becomes
manifest in the $\tau_2\to 0$ region.  
How then can we interpret the increase in the Hagedorn temperature
from the traditional heterotic value  
$T_H=(2-\sqrt{2})\calM$ to 
the new, slightly higher value $T_H=\calM/\sqrt{2}$?
It seems that this would require a corresponding {\it decrease}\/ in the exponential
rate of growth in the asymptotic density of string states. 

To answer this question, we must first recognize that
the transition from the $\calF$-representation to the $\calS$-representation
is highly non-trivial in the case of string theories containing 
spacetime fermions (such as the Type~II and heterotic strings).  
In the case of the bosonic string, for example, the thermal partition function necessarily
takes the factorized form given in Eq.~(\ref{parfunct});  the absence of
spacetime fermions implies that no subsequent thermal $\IZ_2$ orbifolding is required.    
Such a partition function is particularly easy to transform to the $\calS$-representation
where the connection between the degeneracy of states and the Hagedorn temperature
is immediate and apparent in the $\tau_2\to 0$ region
(which is why we do not find a change in the Hagedorn temperature
for bosonic strings, assuming they could be made stable at zero temperature).  
However, as we have seen above,
for Type~II and heterotic strings
the thermal partition function necessarily takes the more complicated
form given in Eq.~(\ref{EOmix}).
The failure of this form to factorize --- indeed, the presence of the half-integer shifts
in the thermal momenta for the $\calE_{1/2}$ and $\calO_{1/2}$ sectors ---  
is the direct consequence of the $\IZ_2$ thermal orbifold. 
However, when we take modular transformations of this partition function
in order to shift to the $\calS$-representation for $\calV(T)$, 
as described in Ref.~\cite{McClainRoth},
this half-integer shift is transformed into non-trivial $\IZ_2$ {\it phases}\/  
(\ie, minus signs) in the corresponding asymptotic degeneracies of states.
These minus signs act to cancel the dominant exponential divergences 
in the degeneracies of states, allowing {\it subleading}\/ exponential divergences
to dominate.  
(Such subleading exponential terms are discussed fully 
in Refs.~\cite{HR,Kani,missusy}.)
This subleading, reduced asymptotic growth 
then correlates directly with the increase in the Hagedorn
temperature that we have found.

Still, one may argue on general conformal-field-theory (CFT) grounds that
such a change in the Hagedorn temperature should not be possible.
After all, the heterotic string has a worldsheet CFT with central charges
$(c_R,c_L)=(12,24)$  in light-cone gauge;  this is implicit 
in the fact that the heterotic string ground state has worldsheet
vacuum energies $(H_R,H_L) = ( -c_R /24,-c_L/24)=(-1/2,-1)$.  The usual
arguments for the Hagedorn transition then lead directly to a Hagedorn temperature
given in terms of these central charges as
\beq
       T_H ~=~   \left(  \sqrt{c_L\over 24} + \sqrt{c_R\over 24}
              \right)^{-1}\, \calM~,
\label{THag}
\eeq
and we have seen that this argument certainly holds in the case of the ten-dimensional
Type~II strings.
However, this argument fails in the case of the 
heterotic strings because the heterotic string ground state, encapsulated
within the CFT character $\chibar_I\chi_I^2$, has been {\it GSO-projected}\/
out of the thermal spectrum in Eq.~(\ref{interp1}).
Indeed, as stressed above, this is the primary difference between
Eq.~(\ref{fake}) and Eq.~(\ref{interp1}).
Or, to phrase this point slightly differently, the GSO-projections inherent in
Eq.~(\ref{interp1}) have deformed the worldsheet CFT of the theory
so that it effectively has a new string ground state consisting
of the $(H_R,H_L)=(-1/2,-1/2)$ tachyons encapsulated within   
the surviving term $\chibar_I (\chi_I\chi_V+\chi_V\chi_I)$
in Eq.~(\ref{interp1}).
Since this effective ground state has depth $(H_R,H_L)=(-1/2,-1/2)$, we can
identify the ``effective'' central charge of this deformed left/right worldsheet theory 
after the GSO projections
as $(c_R,c_L)=(12,12)$, just as for the Type~II string. 

It is also easy to understand how this deformation manages to affect
the asymptotic exponential growth in the degeneracy of states within $Z_{\rm therm}$.  
Recall that in a given CFT with central charge $c$,
each character $\chi_i(\tau)$ 
represents a trace over the Verma module
associated with the corresponding primary field $\phi_i$,
and has a $q$-expansion of the form~\cite{HR,Kani,missusy}
\beq
      \chi_i(\tau) ~=~   q^{h_i-c/24} \, \sum_{p=0}^\infty\,  a_p^{(i)} \, q^p ~~~~~~~~~~~
      \hbox{with}~~
       a_p^{(i)} ~\sim~ \exp\left( 4\pi \sqrt{c p\over 24} \right)~ ~~{\rm as}~ p\to\infty~.
\label{chars}
\eeq
Here $h_i$ is the conformal weight of the primary field $\phi_i$.
In deriving this result via the 
standard contour-integral methods~\cite{HR,Kani,missusy}, 
one uses the fact that each character $\chi_i$
is connected to the identity character $\chi_I$ of the CFT ground state
through a $\tau\to -1/\tau$ modular transformation.
(The existence of this connection is guaranteed from the CFT fusion rules and the 
Verlinde formula~\cite{Verlinde}.)
It is for this reason that the $q$-expansion of each character $\chi_i$ of the CFT {\it individually}\/
has coefficients which
exhibit an exponential growth rate related to the underlying central charge of the
worldsheet CFT.  ~Likewise, this is why {\it individual}\/ products of left- and right-moving 
characters yield asymptotic growth rates consistent with the 
traditional Hagedorn temperature.

However, in our thermal partition function in Eq.~(\ref{interp1}), we  
see that we essentially have four {\it combinations}\/ of left/right characters
which multiply our $\calE/\calO$ functions.  Explicitly, these combinations are given by
\beqn
     Z^{(1)} &=& \chibar_V \,(\chi_I^2 + \chi_V^2)  ~-~  \chibar_S \,(\chi_S^2 + \chi_C^2)\nonumber\\
     Z^{(2)} &=& \chibar_V \,(\chi_S^2+\chi_C^2)  ~-~  \chibar_S \,(\chi_I^2+\chi_V^2) \nonumber\\
     Z^{(3)} &=& \chibar_I \,(\chi_I\chi_V+\chi_V\chi_I) ~-~ \chibar_C \,(\chi_S\chi_C+\chi_C\chi_S) 
                    \nonumber\\
        Z^{(4)} &=& \chibar_I \,(\chi_S\chi_C+\chi_C\chi_S) ~-~ 
           \chibar_C \,(\chi_I\chi_V+\chi_V\chi_I) ~.
\label{newblocks}
\eeqn
These four left/right character
combinations close amongst themselves
under modular transformations, 
and thus function as a new ``effective'' set of characters 
for our ``effective'' (deformed) left/right worldsheet CFT.~  
However, this effective character set does not contain the heterotic CFT 
ground-state $\chibar_I\chi_I^2$;  indeed, the most tachyonic
term that survives in this character set
is the term $\chibar_I (\chi_I\chi_V+\chi_V\chi_I)$ within $Z^{(3)}$.
We thus see that this effective left/right CFT has a reduced ``effective'' central charge
compared with the original left/right CFT prior to GSO projections, as discussed above.  

This is also directly evident from the
$(q,\qbar)$-expansions of the left/right character combinations 
$Z^{(1)}$ and $Z^{(2)}$. 
For example, looking at $Z^{(1)}$, we find that the first term
      $\chibar_V (\chi_I^2 + \chi_V^2)$ individually has
a $(q,\qbar)$-expansion with coefficients (mass-level degeneracies) 
exhibiting an exponential growth rate consistent with the traditional Hagedorn temperature.
However, the same is also true of the second term within $Z^{(1)}$, namely
$\chibar_S (\chi_S^2 + \chi_C^2)$,
and the fact that these terms are {\it subtracted}\/ in forming $Z^{(1)}$
actually ends up {\it cancelling}\/ this leading exponential behavior.
What remains is only a subleading
exponential growth rate consistent with our re-identified Hagedorn temperature.
A similar cancellation also holds for $Z^{(2)}$;
note that cancellations of these sorts are discussed in detail in Ref.~\cite{missusy}.
Thus, at the partition-function level, this reduction in the Hagedorn temperature
is a direct result of the minus signs within the combinations
$Z^{(1)}$ and $Z^{(2)}$
in Eq.~(\ref{newblocks}), which in turn are a direct result of the non-standard
thermal spin-statistics relations
that we have already observed at the Planck scale.

In some sense, this entire argument may be summarized as follows.  Let us look
again at the original partition function of the $SO(32)$ heterotic string
model in Eq.~(\ref{SO32partfunct}).  As a result of spacetime supersymmetry,  
this partition function vanishes identically --- \ie, all of its level-degeneracy coefficients
are identically zero.  There is no exponential growth here at all.
But one does not examine the {\it whole}\/ partition function   
in order to derive a Hagedorn temperature;  one instead looks at its separate bosonic and
fermionic contributions.  Ordinarily, these contributions would
be identified as
\beq
         Z^{\rm (bosonic)}_{SO(32)} = Z^{(8)}_{\rm boson}~
                     \chibar_V \,(\chi_I^2 + \chi_V^2 + \chi_S^2 + \chi_C^2)~,~~~~~~~
         Z^{\rm (fermionic)}_{SO(32)} = Z^{(8)}_{\rm boson}~
                     \chibar_S \,(\chi_I^2 + \chi_V^2 + \chi_S^2 + \chi_C^2)~,
\label{usualcomponents}
\eeq
and indeed
each of these expressions separately exhibits an exponential rise in the degeneracy
of states which is consistent with the traditional heterotic Hagedorn temperature.
But what do we really 
mean by ``bosonic'' and ``fermionic''?
Taking a thermodynamic definition, we are forced to identify such states
according to their periodicities around the thermal circle.
Therefore, given our $(D-1)$-dimensional interpolating-model analysis,
we now see that for the heterotic string,
Eq.~(\ref{usualcomponents}) is not the correct way to separate the total partition function into its
separate thermal components.  Instead, for thermal purposes, we now see that
the proper separation is into the different components 
\beqn
         Z^{\hbox{\small (``bosonic'')}}_{SO(32)} &=& Z^{(8)}_{\rm boson}~
               \left\lbrack
          \chibar_V \,(\chi_I^2 + \chi_V^2)  ~-~  \chibar_S \,(\chi_S^2 + \chi_C^2)\right\rbrack
                                                      ~,\nonumber\\
         Z^{\hbox{\small (``fermionic'')}}_{SO(32)} &=& Z^{(8)}_{\rm boson}~
     \left\lbrack \chibar_S \,(\chi_I^2+\chi_V^2)  ~-~ \chibar_V \,(\chi_S^2+\chi_C^2)\right\rbrack~,
\label{newcomponents}
\eeqn
since these are the components 
that appear in the $\calE_0$ and $\calE_{1/2}$ sectors when the proper thermal extrapolation 
is constructed.
It is therefore {\it these}\/ components which function as the ``bosonic'' and ``fermionic''
contributions as far as thermal effects are concerned, and indeed these are the components
which exhibit the slower exponential growth associated with our re-identified
Hagedorn temperature.
We stress that both Eq.~(\ref{usualcomponents}) and Eq.~(\ref{newcomponents})
correctly separate those massless bosonic and fermionic states which survive
in the field-theory limit.  Their only difference is in their treatment of
the stringy Planck-scale states which have no field-theoretic limits.
In other words, Eq.~(\ref{newcomponents}) correctly identifies bosons and fermions
according to their thermal behaviors;  what is unusual is the connection between
this behavior and the spacetime Lorentz spins 
of the Planck-scale states.

 {\it We thus conclude that all tachyon-free closed strings
in ten dimensions share a universal Hagedorn temperature.
Although the heterotic string would naively appear to have a slightly lower
Hagedorn temperature than the Type~II string
due to its non-level-matched ground state, 
self-consistency also requires a set of non-trivial
GSO projections which
compensate for the non-level-matched ground state
by inducing a cancellation in the asymptotic degeneracies of states,
thereby pushing the associated
Hagedorn temperature back to the Type~II value.}

%=============================================================================
\section{Beyond ten dimensions:~  Additional general observations}
\setcounter{footnote}{0}

In ten dimensions, we found that each of the closed string theories
which are tachyon-free at zero temperature
has a Hagedorn transition associated with a tachyon that emerges in its
thermal extrapolation, with worldsheet
energies $(H_R,H_L)=(-1/2,-1/2)$.
For the supersymmetric Type~II strings, we have seen that this tachyon always 
emerges in the thermal
extrapolation because the only possible $T\to\infty$ endpoints 
for the corresponding interpolating models are the Type~0 strings, 
which necessarily contain these tachyons.
In the case of the $SO(32)$ and $E_8\times E_8$ heterotic strings,
we have seen that we must also
identify the appropriate non-supersymmetric heterotic string
models to serve as suitable $T\to\infty$ endpoints.
While the tachyon-free $SO(16)\times SO(16)$ string could have logically
served as this endpoint, it turns out that this choice would violate 
thermal spin-statistics constraints in both the $SO(32)$ 
and $E_8\times E_8$ cases~\cite{IT,julie}.  
Thus, in each case, our $T\to\infty$ endpoint model 
must be one of the remaining non-supersymmetric string
models [\eg, for the $SO(32)$ string, we found that the endpoint was the 
non-supersymmetric $SO(32)$ string].   
Since each of these remaining non-supersymmetric models has 
tachyons with $(H_R,H_L)=(-1/2,-1/2)$, 
we again have a Hagedorn transition, albeit with a re-identified
Hagedorn temperature.

Given these results, two obvious questions arise.
First, is it a general property that {\it all}\/ heterotic strings 
will have new, re-identified Hagedorn temperatures?
Our belief is that this is indeed the case, regardless of the spacetime
dimension.  As we have argued in Sect.~V, the usual Hagedorn transition
in the heterotic case requires the existence of the term 
$\chibar_I \chi_I^2 \calO_{1/2}$ (or more generally, 
the ground-state character multiplied by $\calO_{1/2}$)
in the corresponding thermal interpolating partition function, yet
the thermally twisted nature of the $\calO_{1/2}$ sector precludes
this from happening.  We believe that this is a general argument which
transcends the particular gravitino-based orbifold argument which
was also provided in Sect.~III. 

A second, equally important issue concerns
whether it might be possible to eliminate the Hagedorn
phase transition completely by finding a zero-temperature 
string model for which the appropriate thermal
extrapolation 
involves a $T\to\infty$ endpoint model which is {\it non-supersymmetric but tachyon-free}\/.
While this did not occur in ten dimensions, this remains a logical possibility
in lower dimensions where many such non-supersymmetric tachyon-free 
string models exist, both of the superstring and heterotic
string variety.

While we do not know the answer to this question in the case of the superstring,
we can prove that it is impossible to evade such a phase transition entirely
in the cases of heterotic strings which are supersymmetric at zero temperature.
Our proof runs as follows~\cite{DL}.
Even if the appropriate $T\to\infty$ endpoint theory happens to be tachyon-free,
there will always exist another state in the thermal extrapolation whose thermal excitations
will trigger a non-trivial phase transition.  This is the so-called ``proto-gravitino''
state:
\beq
        \hbox{proto-gravitino:}~~~~~~~~~~~~
             \tilde \phi^\alpha ~\equiv~
         \lbrace \tilde b_{0}\rbrace^\alpha  |0\rangle_R ~\otimes~ ~ |0\rangle_L~.
\label{protogravitino}
\eeq
Note that this state is constructed exactly as the gravitino in Eq.~(\ref{gravitino}),
but without the left-moving worldsheet coordinate excitation.
However, it is important to realize
that {\it GSO projections are completely insensitive to
the presence or absence of excitations of the worldsheet coordinate
bosonic fields}\/.  Thus, if our zero-temperature heterotic string model is 
supersymmetric and the gravitino
is therefore present in the original zero-temperature theory, then 
the proto-gravitino must also always be present in the original zero-temperature theory.
           
Since this state emerges, like the gravitino itself,
from the untwisted (fermionic) gravity sector of the original $T\to 0$ model, 
its contributions must appear multiplied by $\calE_{1/2}$ within any self-consistent 
heterotic thermal extrapolation away from that model.  
[For example, in the case of the $SO(32)$ heterotic
string interpolation in Eq.~(\ref{interp1}), 
the proto-gravitino contribution was hiding within the term $-\chibar_S \chi_I^2 \calE_{1/2}$.]
However, this proto-gravitino state is then necessarily endowed with a thermal $(m,n)=(1/2,2)$
excitation which is massive for all $T\not= T_\ast$ but
exactly massless at the single temperature $T=T_\ast$ where $T_\ast\equiv 2\calM$.
(Note that since this state is fermionic, Lorentz invariance prevents it from
becoming tachyonic at any temperature.)
The sudden appearance of a new massless state at $T=T_\ast$ signals
the emergence of long-range order in the thermal theory, and can again be associated
with a Hagedorn-like phase transition.  However, since this state hits
masslessness only once as a function of temperature and never becomes tachyonic,
this turns out~\cite{DL} to be a very weak, $p^{\rm th}$-order phase transition,
where $p$ is related to the spacetime dimension $D$:
\beq
            p ~=~ \cases{  D & for $D$ even \cr
                           D-1 & for $D$ odd~.\cr}
\label{order}
\eeq

Thus, we conclude that for supersymmetric heterotic strings,
it is never possible to completely evade a Hagedorn-like phase transition.
However, the phase transition associated with the proto-gravitino
state appears only at the relatively high temperature $T_\ast\equiv 2\calM$,
and thus will be completely irrelevant if tachyon-induced Hagedorn
transitions appear at lower temperatures.

%=============================================================================
\section{Conclusions}
\setcounter{footnote}{0}

In this paper, we investigated the manner in which a given zero-temperature
string model may be extrapolated to finite temperature.
Following relatively conservative conditions for self-consistency,
we nevertheless found that the traditional Hagedorn transition does
not exist for heterotic strings but is instead replaced by a new, ``re-identified''
Hagedorn transition which emerges at the somewhat higher temperature 
normally associated with Type~II strings.  
This allowed us to uncover a universal Hagedorn temperature 
for all tachyon-free closed string theories in ten dimensions.  
We also showed that these results are not in conflict with 
the exponential rise in the degeneracy of string states in these models.

Clearly, many outstanding questions remain.
Perhaps the two most critical are the issues of the {\it existence}\/ and {\it uniqueness}\/
of thermal extrapolations satisfying the general criteria we put forth in Sect.~II.~  
In other words, it is important to demonstrate that, for any given $D$-dimensional
zero-temperature string model,
there always exists one and only one suitable corresponding $T\to\infty$ endpoint $D$-dimensional
string model such
that the corresponding $(D-1)$-dimensional interpolation is thermally consistent according
to our general criteria, including proper spin-statistics relations.
In ten dimensions, we have already seen that such extrapolations exist and are unique.
However, neither property has been proven in lower dimensions.  
This is clearly an important  
issue that requires further study.

Another interesting question concerns the thermal fate of string models which
are non-supersymmetric but tachyon-free at zero temperature:  is it ever possible
that such a non-supersymmetric model will have a thermal extrapolation whose $T\to\infty$ limit
is {\it supersymmetric}\/?  If so, this would be an example of a situation in which
the zero-temperature theory is non-supersymmetric,
but in which thermal effects compensate for this inequity between bosons
and fermions and thereby {\it introduce}\/ (rather than break) supersymmetry as $T\to\infty$.
In other words, such thermal effects would be ``SUSY-making'' rather than SUSY-breaking,
with SUSY-breaking occurring at {\it lower}\/ temperatures.
This phenomenon would be intrinsically string-theoretic, since only for
closed strings does the $T\to\infty$ limit yield a theory of the same dimensionality  
as the original zero-temperature theory. 
No examples exhibiting this phenomenon exist in ten dimensions, but it would
be interesting to explore whether such examples might exist in lower dimensions.

%=============================================================================
\section*{Acknowledgments}
\setcounter{footnote}{0}

This work is supported in part by the
National Science Foundation
under Grant~PHY/0301998,
by the Department of Energy under Grant~DE-FG02-04ER-41298,
and by a Research Innovation Award from
Research Corporation.
We are happy to thank
J.~Davis, E.~Dudas, J.~Erlich, E.~Kiritsis,
F.~Larsen, L.~Motl, and G.~Shiu for discussions.
One of us (KRD) also wishes to thank the Aspen Center for Physics
for hospitality while this paper was written.

%  \vfill\eject
\bigskip
\bigskip
\bigskip

%=============================================================================
\leftline{\bf Note Added:}

More than three weeks after these results originally appeared in Ref.~\cite{DL},
another article appeared~\cite{Shyamoli}
whose author claims that there is no Hagedorn transition whatsoever
for bosonic strings, Type~II strings, heterotic strings, or even Type~I strings.
Clearly, 
we disagree with the results of that paper, and shall limit our comments
to version~4 of Ref.~\cite{Shyamoli} (which is the current version as we write this).  
Although the modular-invariance errors of the previous versions have been corrected,  
we believe that the relevant 
thermal expressions given in version~4
for Type~II and heterotic
strings (Eqs.~(23) and (67) of Ref.~\cite{Shyamoli}, respectively)
are still manifestly inconsistent.   
One error, for example, concerns thermal spin-statistics:
both expressions contain massless bosonic and fermionic
states with incorrect modings around the thermal circle.
[For example, the first line 
of Eq.~(23) of Ref.~\cite{Shyamoli}  necessarily contains contributions from
the gravity multiplet, yet these are multiplied by the {\it sum}\/ $\calE_0+\calE_{1/2}$, 
thereby giving the graviton both periodic {\it and}\/ anti-periodic thermal excitations
around the thermal circle!]
Second, Eq.~(23) of Ref.~\cite{Shyamoli} does not represent a self-consistent interpolation,
since the $T\to \infty$ limit does not correspond to any ten-dimensional Type~II string.
Third, although the author of Ref.~\cite{Shyamoli} claims to find 
an {\it infinite}\/ number of thermal tachyons  in the Type~II thermal theory starting at $T=0$, 
this is because the $T\to 0$ limit of Eq.~(23) of Ref.~\cite{Shyamoli} is 
actually the tachyonic Type~0A/0B theory, not the desired Type~IIA/B theory.
Indeed, although the author claims that Eq.~(23) of Ref.~\cite{Shyamoli}
is the {\it unique}\/ modular-invariant
Type~II thermal partition function, the correct partition function is 
actually our Eq.~(\ref{TypeIIinterp}), which leads to the usual 
Hagedorn transition, as expected.  
Finally, the author of Ref.~\cite{Shyamoli} claims to see no evidence for the
heterotic phase transition that we discussed
above, induced by the proto-gravitino state.  Once again, this is due to the inconsistency
of the expressions in Ref.~\cite{Shyamoli};  as we proved in our original paper~\cite{DL}, 
this state must always exist if the $T\to 0$ heterotic theory is supersymmetric.
Indeed, when we fix Eq.~(67) of Ref.~\cite{Shyamoli} by 
copying the results from previous papers~\cite{IT}
correctly and with the correct {\it supersymmetric}\/ $T\to 0$ limit,  
we find exactly the behavior that we described above.
Indeed, this sort of phase transition has been recently discussed by other authors as well
in other contexts (see, \eg, Ref.~\cite{2Dhet}).
Overall, however, the biggest difference between
our approach and that of Ref.~\cite{Shyamoli} is that we construct
our thermal extrapolations by demanding
the physical criteria as we laid out in our Sect.~II;
we then determine the properties of the Hagedorn transition 
as a consequence.  By contrast,
the author of Ref.~\cite{Shyamoli} is implicitly requiring that there be no Hagedorn transition
as a {\it precondition}\/ when demanding that all thermal extrapolations 
be tachyon-free at all temperatures, regardless of other self-consistency checks.

The purpose of this Note Added has merely been to clarify the primary differences between
our work and that of Ref.~\cite{Shyamoli}, as these differences have 
apparently confused several readers.

\bigskip
\bigskip
\bigskip

%=============== REFERENCES HERE ================

%================================================


\begin{references}




\bibitem{Polbook}
      For an introduction, see:\\
        M.~B.~Green, J.~A.~Schwarz and E.~Witten,
        {\it Superstring Theory, Vols.~I and II}\/
        (Cambridge University Press, 1987);\\
      J. Polchinski, {\it String Theory, Vols.~I and II}\/
      (Cambridge University Press, 1998).  %  Chap.~9.


\bibitem{Pol86}   J.~Polchinski,
     %``Evaluation Of The One Loop String Path Integral,''
     Commun.\ Math.\ Phys.\  {\bf 104}, 37 (1986).
     %%CITATION = CMPHA,104,37;%%

\bibitem{Rohm}
      R.~Rohm,
      %``Spontaneous Supersymmetry Breaking In Supersymmetric String Theories,''
      Nucl.\ Phys.\ B {\bf 237}, 553 (1984).
      %%CITATION = NUPHA,B237,553;%%


\bibitem{AtickWitten}
     J.~J.~Atick and E.~Witten,
     %``The Hagedorn Transition And The Number Of Degrees Of Freedom Of String Theory,''
     Nucl.\ Phys.\ B {\bf 310}, 291 (1988).
     %%CITATION = NUPHA,B310,291;%%


\bibitem{KLTclassification}
  H.~Kawai, D.~C.~Lewellen and S.~H.~H.~Tye,
  %``Classification Of Closed Fermionic String Models,''
  Phys.\ Rev.\ D {\bf 34}, 3794 (1986).
  %%CITATION = PHRVA,D34,3794;%%

\bibitem{julie}
     J.~D.~Blum and K.~R.~Dienes,
     %``Duality without supersymmetry: The case of the SO(16) x SO(16) string,''
     Phys.\ Lett.\ B {\bf 414}, 260 (1997)
     [arXiv:hep-th/9707148];
     %%CITATION = HEP-TH 9707148;%%
     %``Strong/weak coupling duality relations for non-supersymmetric string
     %theories,''
     Nucl.\ Phys.\ B {\bf 516}, 83 (1998)
     [arXiv:hep-th/9707160].
     %%CITATION = HEP-TH 9707160;%%


\bibitem{Hagedorn}
  R.~Hagedorn,
  %``Statistical Thermodynamics Of Strong Interactions At High-Energies,''
  Nuovo Cim.\ Suppl.\  {\bf 3}, 147 (1965).
  %%CITATION = NUCUA,3,147;%%


\bibitem{DL}  K.~R.~Dienes and M.~Lennek,
  %``Re-identifying the Hagedorn transition,''
  arXiv:hep-th/0505233.
  %%CITATION = HEP-TH 0505233;%%

\bibitem{McClainRoth}
  B.~McClain and B.~D.~B.~Roth,
  %``Modular Invariance For Interacting Bosonic Strings At Finite Temperature,''
  Commun.\ Math.\ Phys.\  {\bf 111}, 539 (1987);\\
  %%CITATION = CMPHA,111,539;%%
  K.~H.~O'Brien and C.~I.~Tan,
  %``Modular Invariance Of Thermopartition Function And Global Phase Structure
  %Of Heterotic String,''
  Phys.\ Rev.\ D {\bf 36}, 1184 (1987).
  %%CITATION = PHRVA,D36,1184;%%

\bibitem{HR}
  G.~H.~Hardy and S.~Ramanujan,
  Proc. London Math. Soc. {\bf 17}, 75 (1918).

\bibitem{Kani}
  I.~Kani and C.~Vafa,
  %``Asymptotic Mass Degeneracies In Conformal Field Theories,''
  Commun.\ Math.\ Phys.\  {\bf 130}, 529 (1990).
  %%CITATION = CMPHA,130,529;%%


\bibitem{missusy}
  K.~R.~Dienes,
  %``Modular invariance, finiteness, and misaligned supersymmetry: New
  %constraints on the numbers of physical string states,''
  Nucl.\ Phys.\ B {\bf 429}, 533 (1994)
  [arXiv:hep-th/9402006].


\bibitem{Verlinde}
  E.~P.~Verlinde,
  %``Fusion Rules And Modular Transformations In 2-D Conformal Field Theory,''
  Nucl.\ Phys.\ B {\bf 300}, 360 (1988);\\
  %%CITATION = NUPHA,B300,360;%%
  G.~W.~Moore and N.~Seiberg,
  %``Polynomial Equations For Rational Conformal Field Theories,''
  Phys.\ Lett.\ B {\bf 212}, 451 (1988).
  %%CITATION = PHLTA,B212,451;%%

\bibitem{IT}
     H.~Itoyama and T.~R.~Taylor,
      %``Supersymmetry Restoration In The Compactified O(16) X O(16)-Prime Heterotic
     %String Theory,''
     Phys.\ Lett.\ B {\bf 186}, 129 (1987).
     %%CITATION = PHLTA,B186,129;%%


\bibitem{Shyamoli}
    S.~Chaudhuri,
    %``Dispelling the Hagedorn myth: Canonical and microcanonical strings,''
    arXiv:hep-th/0506143.
    %%CITATION = HEP-TH 0506143;%%


\bibitem{2Dhet}
  J.~L.~Davis, F.~Larsen and N.~Seiberg,
  %``Heterotic strings in two dimensions and new stringy phase transitions,''
  arXiv:hep-th/0505081.
  %%CITATION = HEP-TH 0505081;%%





\end{references}
\end{document}